\documentclass[pre,eqsecnum,aps]{revtex4-2}

\usepackage{physics}
\usepackage{float}

\usepackage{amsmath}
\usepackage{comment}
\usepackage{graphicx}

\usepackage{xcolor}
\definecolor{apsblue}{rgb}{0.18,0.19,0.57} % Standard APS/PRE dark blue

\usepackage{url} % Fixes the underscore issue in URLs
\usepackage[
    colorlinks=true, 
    citecolor=apsblue, 
    urlcolor=apsblue, 
    linkcolor=apsblue
]{hyperref}

\begin{document}

\title{Newton's identity in finite-bead fermionic partition function}

\author{A. Chaudhary}
\email{aarifchaudharyg@gmail.com}
\author{J. Valenzuela}
\email{jonasvalt@tamu.edu}
\affiliation{Department of Physics and Astronomy, Texas A\&M University, College Station, TX 77843, USA}

\begin{abstract}
For non-interacting fermions in a harmonic trap, the partition function at any discrete number of imaginary time slices (or beads) and for any choice of short-time propagator admits an exact recursion relation derived directly from the contracted determinant form of the path integral. This finite-bead recursion is distinct from earlier continuum-limit recursions, which do not apply to the discrete time partition functions. By identifying a direct correspondence between this recursion and Newton's identity, application of a closed-form result from the theory of partitions provides an exact expression for the one-dimensional $n$-fermion finite-bead partition function. From this, the thermodynamic and Hamiltonian energies and specific heats are analytically calculated for any $n$, $N$, $\tau$, and propagator choice.
\end{abstract}

\maketitle
\section{Introduction}

While the origin of the sign problem in Path Integral Monte Carlo (PIMC) due to fermionic antisymmetry is well understood, explicit analytical study of finite imaginary time has remained limited. In this work, we lay the analytical groundwork for the study of the fermionic sign problem at finite imaginary time. We consider $n$ non-interacting fermions in a $1$-dimensional harmonic trap. Starting from the determinant form of the fermionic propagator, all intermediate position integrations can be performed in closed form\cite{chin2026understandingsignproblemexact}. This yields an exact recursion relation for the partition function in terms of the finite-bead cycle trace, valid for any number of particles $n$, number of beads $N$, and short-time propagator choice.

Recursion relations of this type have a long history, but prior derivations apply only to the exact continuum partition function rather than to a finite number of beads. Ford\cite{Ford71} first observed a correspondence between the symmetric functions of statistical mechanics and the non-interacting partition function; however, this connection was heuristic and was neither derived nor exploited to produce a closed form. Borrmann and Franke\cite{10.1063/1.464180} subsequently showed that the same recursion follows directly from the permutation (cycle) decomposition of the partition function, a purely combinatorial argument that makes no reference to a path integral. Brosens, Lemmens, and Devreese\cite{PhysRevE.55.227} then derived the recursion from the path-integral definition itself in the continuum limit, again through the permutation decomposition, and for the harmonic oscillator reported a closed-form partition function in one-dimension. However, that closed form was obtained by verification rather than derivation: it can be checked to satisfy the recursion, but no constructive route to it was given.

The present work differs from all of these by deriving the {\it finite-bead} recursion directly from the determinant form rather than permutation expansion. This recursion is again an instance of Newton's identity relating power sums to elementary symmetric polynomials, but the power sums are now finite-bead cycle traces — products of the short-time propagator over a finite number of beads — rather than continuum single-particle partition functions. They depend explicitly on $N$, on $\tau$, and on the chosen short-time approximation, so the recursion describes the discretized path integral that PIMC actually samples at any finite bead number, not merely its continuum limit. %Since it descends from the determinant propagator rather than a permutation expansion, identifying it as Newton's identity makes rigorous the symmetric-function correspondence that Ford \cite{Ford71} observed only heuristically.

Because the inputs to Newton's identity differ, so does its solution. For the one-dimensional harmonic oscillator the finite-bead cycle traces admit a geometric-series expansion that makes the underlying variables explicit, and a classical result from Macdonald\cite{macdonald1998symmetric} yields a closed form for the $n$-fermion, $N$-bead partition function. This is a different function from the continuum expression of Brosens et al.\cite{PhysRevE.55.227}: it depends on $N$ and the short-time propagator, recovers their result only as $N\to\infty$, and is derived rather than verified after the fact, supplying the construction their work left open.

From this exact partition function, the Thermodynamic and Hamiltonian energies, as well as the specific heats, are calculated analytically. These results appropriately recover the $\tau \to \infty$ continuum-limit fermion energy $E_n = n^2/2$ and reproduce the small $n$ calculations of Takahashi and Imada\cite{tak84a}. Because the sign problem is absent for one-dimensional configurations\cite{tak84a,PhysRevE.107.035305} the $d=1$ results do not tackle the sign problem directly; what they provide instead is a setting in which every quantity is known exactly at every bead number making the formulation a rigorous diagnostic baseline for PIMC simulation. This formulation also serves as a direct foundation for an extension to the $d=2$ case, where the sign problem is present (see end of Section \ref{rfpf}).

\section{Short-time propagators}
\label{shp}

Consider the one-dimensional harmonic oscillator Hamiltonian
\begin{equation*}
    \hat{H} = -\frac{1}{2}\frac{d^2}{dx^2} + \frac{1}{2}x^2 = \hat{T} + \hat{V}.
\end{equation*}
The 1-bead short-time approximation to the imaginary-time propagator $\langle x' | e^{-\epsilon(\hat{T} + \hat{V})} |
x \rangle$, from Ref.\cite{10.1063/5.0164086} is 
\begin{equation*}
     G_1(x', x, \epsilon) = \frac{1}{\sqrt{2\pi\kappa_1(\epsilon)}} e^{-\mu_1(\epsilon)\frac{1}{2}x'^2} e^{-\frac{1}{2\kappa_1(\epsilon)}(x' - x)^2} e^{-\mu_1(\epsilon)\frac{1}{2}x^2}.
\end{equation*}
For the second-order primitive approximation (PA) propagator,
\begin{equation}
     \kappa_1(\epsilon) = \epsilon, \quad \mu_1(\epsilon) = \frac{\epsilon}{2}, \quad \zeta_1 = 1 + \frac{\epsilon^2}{2}, \quad \gamma = \sqrt{1 + \frac{\epsilon^2}{4}}.
\label{algpa}
\end{equation}
For the fourth order 4A propagator (4A)\cite{10.1063/1.1485725},
\begin{align}
\kappa_1(\epsilon) &= \epsilon(1 + \frac{\epsilon^2}{12})^2, \quad \mu_1(\epsilon) = \left(\frac{\epsilon}{2} + \frac{\epsilon^3}{24} + \frac{\epsilon^5}{864}\right) \bigg/ \left(1 + \frac{\epsilon^2}{12}\right)^2,\nonumber\\ 
\zeta_1(\epsilon) &= 1 + \frac{\epsilon^2}{2} + \frac{\epsilon^4}{24} + \frac{\epsilon^6}{864}, \quad
\gamma = \sqrt{1 + \epsilon^4 / (432 + 36\epsilon^2)}.
\end{align}
where $\zeta_1=1+\kappa_1\mu_1$ and $\gamma=\sqrt{\zeta_1^2-1}/\kappa_1=\sqrt{\zeta_N^2-1}/\kappa_N$ are needed below to compute the Thermodynamic and Hamiltonian energies.
The approximate $N$-bead propagator is then given by:
\begin{equation}
\label{eq:contraction}
\begin{aligned}
G_N(x', x, \tau) &= \langle x' |
(e^{-\epsilon(\hat{T} + \hat{V})})^N | x \rangle \\
&= \int_{-\infty}^{\infty} dx_1 \cdots dx_{N-1} G_1(x', x_1, \epsilon) G_1(x_1, x_2, \epsilon) \cdots G_1(x_{N-1}, x, \epsilon)
\\
&= \frac{1}{\sqrt{2\pi\kappa_N(\tau)}}e^{-\frac{1}{2}\mu_N(\tau)(x'^2+x^2)}e^{-\frac{1}{2\kappa_N(\tau)}(x'-x)^2},
\end{aligned}
\end{equation}
where $\tau = N\epsilon$ and $\kappa_N(\tau)$ and $\mu_N(\tau)$ are coefficients after contracting\cite{10.1063/5.0164086} $N$ short-time propagators above.
Since the potential is non-interacting, the fermion determinant propagator is just a permutation sum of bosonic propagators and hence the contraction holds for each permutation which recombines to give a single determinant.
The $N$-bead partition function then follows from
\begin{equation*}
    Z^N(\tau) = \int dx G_N(x,x,\tau).
\end{equation*}
Our focus is to derive $Z^N(\tau)$ analytically, from which the energy can be computed.

\section{Recursive finite-bead fermion partition function}
\label{rfpf}

The propagator for $n$ fermions admits a contraction as in Eq. (\ref{eq:contraction}) with a single determinant, valid for any amount of beads $N$ and propagator of choice \cite{chin2026understandingsignproblemexact}. In one-dimension, it can be written as
\begin{equation*}
    G^N_n(x',x,\tau) = \frac{1}{\sqrt{(2\pi\kappa_N(\tau))^n}} \prod_{i=1}^{n} e^{-\frac{1}{2}\mu_N(\tau)({x'}_i^2+x_i^2)} \det(K) \,,\,\, K_{ij}(x',x) = e^{-\frac{1}{2\kappa_N(\tau)}(x_i'-x_j)^2},
\end{equation*}
where $x = (x_1, \dots, x_n)$ and likewise for $x'$.
The corresponding $n$-fermion partition function is given by
\begin{equation*}
    Z_n^N(\tau) = \frac{1}{n!}\frac{1}{\sqrt{(2\pi\kappa_N(\tau))^n}}\prod_{i=1}^{n} \int dx_i e^{-\mu_N(\tau) x_i^2} \det(K), \quad K_{ij}(x) = e^{-\frac{1}{2\kappa_N(\tau)}(x_i - x_j)^2}.
\end{equation*}
Since
\[
K = \begin{pmatrix}
1 & \xi_{1,2} & \cdots & \xi_{1,n} \\
\xi_{2,1} & 1 & \cdots & \xi_{2,n} \\
\vdots & \vdots & \ddots & \vdots \\
\xi_{n,1} & \xi_{n,2} & \cdots & 1
\end{pmatrix},
\quad
\xi_{i,j} = e^{-\frac{1}{2\kappa_N(\tau)}(x_i - x_j)^2},
\]
the first term of $\det(K)$ is simply $1 \, |K_{n-1}|$, where $|K_{n-1}|$ is the $(n-1)\times (n-1)$ minor obtained by removing row 1 and column 1. The integral then simplifies to:
\begin{equation}
\begin{aligned}
     &\frac{1}{\sqrt{(2\pi\kappa_N(\tau))^n}}\prod_{i=1}^{n} \int dx_i e^{-\mu_N(\tau) x_i^2} |K_{n-1}|
\\ 
     &= \frac{1}{\sqrt{(2\pi\kappa_N(\tau))^n}}\int dx_1 e^{-\mu_N(\tau) x_1^2} \prod_{i=2}^{n} \int dx_i e^{-\mu_N(\tau) x_i^2} |K_{n-1}|
\\  
     &= z_1^N  Z_{n-1}^N.
\end{aligned}
\label{eq:SepIng}
\end{equation}
For brevity, the prefactor $1/n!$ is temporarily omitted and will be restored at the end of the derivation.
Expanding the determinant along the first row, the second term is given by
\begin{equation}
\label{term2}
-\xi_{1,2}
\begin{vmatrix}
\xi_{2,1} & \xi_{2,3} & \cdots & \xi_{2,n} \\
\xi_{3,1} & 1       & \cdots & \xi_{3,n} \\
\vdots  & \vdots  & \ddots & \vdots  \\
\xi_{n,1} & \xi_{n,3} & \cdots & 1 
\end{vmatrix},
\end{equation}
and the $i$-th term takes the form
\begin{equation*}
(-1)^{i+1}\xi_{1,i}
\begin{vmatrix}
\xi_{2,1} & 1       & \cdots & \xi_{2,i-1}  & \xi_{2,i+1} & \cdots & \xi_{2,n}\\
\vdots  & \vdots  & \vdots & \vdots     & \vdots    & \vdots & \vdots \\
\xi_{i,1} & \xi_{i,2} & 
\vdots & \text{all }\xi\text{ row} & \vdots    & \vdots & \vdots \\
\vdots  & \vdots  & \vdots & \vdots     & \vdots    & \vdots & \vdots \\
\xi_{n,1} & \xi_{n,3} & \cdots & \cdots     & \cdots    & \cdots & 1 \\
\end{vmatrix},
\end{equation*}
where the row denoted ``all $\xi$ row'' consists entirely of off-diagonal Gaussian factors and lacks a diagonal element of $1$.
By sequentially exchanging this $i$-th row with the preceding row until it reaches the top of the minor, we perform $i-2$ row swaps.
Each swap introduces a negative sign, yielding an overall parity factor of $(-1)^{(i+1)+(i-2)} = (-1)^{2i-1} = -1$.
The $i$-th term thus becomes:
\begin{equation*}
-\xi_{1,i}
\begin{vmatrix}
\xi_{i,1} & \xi_{i,2} & \cdots & \xi_{i,i-1}  & \xi_{i,i+1} & \cdots & \xi_{i,n}\\
\xi_{2,1} & 1       & \cdots & \cdots     & \cdots    & \cdots & \vdots \\
\vdots  & \vdots  & \ddots & \cdots     & \cdots    & \cdots & \vdots \\
\xi_{i-1,1} & \xi_{i-1,2} &\vdots & \ddots    & \cdots & \cdots & \vdots \\
\xi_{i+1,1} & \xi_{i+1,2} &\vdots & \vdots    & \ddots & \cdots & \vdots \\
\vdots  & \vdots  
\vdots & \vdots     & \vdots    & \ddots & \vdots & \vdots \\
\xi_{n,1} & \xi_{n,3} & \cdots & \cdots     & \cdots    & \cdots & 1 \\
\end{vmatrix}.
\end{equation*}
Because the variables are dummy indices of integration, they can be cyclically relabeled ($2 \to 3, 3 \to 4, \dots, i \to 2$).
After this substitution, we obtain:
\begin{equation*}
-\xi_{1,2}
\begin{vmatrix}
\xi_{2,1} & \xi_{2,3} & \cdots & \xi_{2,i}  & \xi_{2,i+1} & \cdots & \xi_{2,n}\\
\xi_{3,1} & 1       & \cdots & \cdots     & \cdots    & \cdots & \vdots \\
\vdots  & \vdots  & \ddots & \cdots     & \cdots    & \cdots & \vdots \\
\xi_{i,1} & \xi_{i,3} &\vdots & \ddots    & \cdots & \cdots & \vdots \\
\xi_{i+1,1} & \xi_{i+1,3} &\vdots & \vdots    & \ddots & \cdots & \vdots \\
\vdots  & \vdots  
\vdots & \vdots     & \vdots    & \ddots & \vdots & \vdots \\
\xi_{n,1} & \xi_{n,3} & \cdots & \cdots     & \cdots    & \cdots & 1 \\
\end{vmatrix}
= -\xi_{1,2}
\begin{vmatrix}
\xi_{2,1} & \xi_{2,3} & \cdots & \xi_{2,n} \\
\xi_{3,1} & 1       & \cdots & \xi_{3,n} \\
\vdots  & \vdots  & \ddots & \vdots  \\
\xi_{n,1} & \xi_{n,3} & \cdots & 1 
\end{vmatrix}.
\end{equation*}
We observe that the $i$-th term contributes equivalently to the second term.
Consequently, there are $(n-1)$ identical contributions equivalent to Eq. (\ref{term2}).
The partition function then reads:
\begin{equation}
    Z_n^N = z_1^N Z_{n-1}^N- (n-1)\prod_{i=1}^{n} \int dx_i e^{-\mu_N(\tau) x_i^2} 
    \xi_{1,2}
    \begin{vmatrix}
    \xi_{2,1} & \xi_{2,3} & \cdots & \xi_{2,n} \\
    \xi_{3,1} & 1       & \cdots & \xi_{3,n} \\
    \vdots  & \vdots  & \ddots & \vdots  \\
    \xi_{n,1} & \xi_{n,3} & \cdots & 1 
    \end{vmatrix}.
\label{eq:SecDet}
\end{equation}
Examining the first term of the inner determinant yields $\xi_{1,2}^2 |K_{n-2}|$. Integrating this component following the same method as in Eq. (\ref{eq:SepIng}) shows that it corresponds to $z_2^N Z_{n-2}^N$. Applying this expansion recursively to the remaining terms of the inner determinant yields a structure similar to Eq. (\ref{eq:SecDet}):
\begin{equation*}
    Z_n^N = z_1^N Z_{n-1}^N - (n-1)\left(z_2^N Z_{n-2}^N -(n-2)\prod_{i=1}^{n} \int dx_i e^{-\mu_N(\tau) x_i^2} 
    \xi_{1,2} \xi_{2,3}
    \begin{vmatrix}
    \xi_{3,1} & \xi_{3,4} & \cdots & \xi_{3,n} \\
    \xi_{4,1} & 1       & \cdots & \xi_{4,n} \\
    \vdots  & \vdots  & \ddots & \vdots  \\
    \xi_{n,1} & \xi_{n,4} & \cdots & 1 
    \end{vmatrix}
    \right).
\end{equation*}
Iterating this expansion systematically yields the following general form
\begin{equation}
\begin{split}
    Z_n^N
    &=z_1^N Z_{n-1}^N - (n-1)\left(z_2^N Z_{n-2}^N -(n-2)\left(z_3^N Z_{n-3}^N -(n-3) \left( \cdots \left( z_{n-1}^N Z_1^N - z_n^N
    \right) \cdots \right) \right) \right) \\ \\
    &= z_1^N 
Z_{n-1}^N -(n-1)z_2^N Z_{n-2}^N + (n-1)(n-2) z_3^N Z_{n-3}^N + \dots + (-1)^{n-1}(n-1)!z_n^N \\ \\
    &= \sum_{i=1}^n (-1)^{i-1} \frac{(n-1)!}{(n-i)!} z_i^N Z_{n-i}^N \,\,\,\, , Z_0^N(\tau) = 1.
\end{split}
\end{equation}
Restoring the $1/n!$ prefactor to the definition of $Z_n^N$ on both sides yields the final recursion relation:
\begin{equation}
    Z_n^N = \frac{1}{n} \sum_{i=1}^n (-1)^{i-1} z_i^N Z_{n-i}^N.
\label{eq:Zrecur}
\end{equation}
Crucially, this recursion holds for any finite number of time slices (beads) $N$ and for any choice of short-time propagator. It is therefore a finite-bead recursion relation, distinct from earlier continuum formulations\cite{Ford71,10.1063/1.464180,PhysRevE.55.227,Schmidt_2002}, which apply only to the exact continuum partition function. Using this recursion, $Z_n^N$ can be determined from the finite-bead cycle traces $z_i^N$.
As detailed in Appendix \ref{App:Tridiag}, evaluating the $i$-cycle traces yields
\begin{equation*}
    z_i^N = \frac{1}{2\sinh(iNu/2)},
\end{equation*}
where $u = \ln(\zeta_1 + \sqrt{\zeta_1^2 - 1})$.
Expressing this result in terms of the parameter $b = e^{-Nu}$, we obtain:
\begin{equation}
    z_i^N = \frac{b^{i/2}}{1-b^{i}}.
\label{eq:smallEres}
\end{equation}
We again emphasize that this cycle trace result holds for any finite-bead propagator, giving a finite $N$ expression distinct from the previous continuum-limit results\cite{PhysRevE.55.227}.

Since the harmonic oscillator potential is separable across spatial dimensions, the $d$-dimensional partition function follows a parallel structure.
The determinant analysis is identical to the $d=1$ case, and the derivation proceeds component-wise, leading to the generalized recursion:
\begin{equation}
\begin{split}
    Z_n^{N,d} &= \frac{1}{n} \sum_{i=1}^n (-1)^{i-1} (z_i^N)^d Z_{n-i}^{N,d},\\
    (z_i^N)^d &= \left(\frac{b^{i/2}}{1-b^{i}}\right)^d.
\end{split}
\label{mainRecursion}
\end{equation}
Solving this general recursive relation is much more involved, and hence in this work we only focus on solving the $d=1$ case, and hence we drop the $d$ superscript. Furthermore, this recursion also holds at the integrand level for arbitrary interactions, agreeing with the determinant after integration (Appendix~\ref{App:Interactions}).

\section{Newton's identity and closed form of $Z_n^N$}

In order to obtain a closed-form solution of $Z_n$, a brief overview of elementary symmetric polynomials (ESPs) as well as Newton's identity is provided. ESPs are defined as the coefficients of powers of $x$, when the linear factorization of a single-variable polynomial is expanded, i.e.,
\begin{equation*}
    \sum_{k = 0}^m(-1)^k e_k x^{m-k} = \prod_{k=1}^m(x-x_k).
\end{equation*}
The ESPs obey the following identity, known as Newton's identity.
\begin{equation}
    e_k = \frac{1}{k} \sum_{i=1}^k (-1)^{i-1} e_{k-i} p_i.
    \label{eq:Zbute}
\end{equation}
Here $p_i$ is the power sum of $(x_1,x_2,...,x_m)$,
\begin{equation*}
    p_i = \sum_{j=1}^m x_j^i.
\end{equation*}
Given a closed form for $p_r$ of infinite roots, Macdonald provides the following closed form result for $e_r$\cite{macdonald1998symmetric}:
\begin{equation}
\begin{split}
    e_r = \prod_{i=1}^r \frac{aq^{i-1}-c}{1-q^i} \\
    p_r = (a^r-c^r)/(1-q^r).
\end{split}   
     \label{eq:BookEq}   
\end{equation}

To solve the recursive relation for $Z_n^N$, we map the ESPs and power sums to $Z_k^N$ and $z_i^N$, respectively (i.e. $e_k = Z_k^N$ and $p_i = z_i^N$).
The structural equivalence of this mapping can be verified by comparing Eq. (\ref{eq:Zbute}) with Eq. (\ref{eq:Zrecur}).
To obtain the corresponding mapping for the roots $x_k$, we Taylor expand Eq. (\ref{eq:smallEres}):
\begin{equation*}
    z_i^N = b^{i/2}\sum_{k=0}^\infty(b^i)^k = \sum_{k=1}^\infty(b^{k-1/2})^i
    = p_i = \sum_{k=1}^\infty x_k^i.
\end{equation*}
This implies the existence of infinitely many roots given by $x_k = b^{k-1/2}$, for which the power series converge since $0<b<1$. Using this one-to-one mapping, we can apply Eq. (\ref{eq:BookEq}) with $a = b^{1/2}$, $c = 0$, and $q = b$ to obtain a closed form for $Z_n^N$,
\begin{equation}
    Z_n^N = b^{n^2/2}\prod_{i=1}^n \frac{1}{1-b^i}.
    \label{eq:ClosedZ}
\end{equation}
In the continuum limit, Eq. (\ref{eq:ClosedZ}) reproduces the partition function obtained by Brosens et al.\cite{PhysRevE.55.227}, which was originally derived independent of the recursion relation.

Beyond yielding the closed form, identifying the recursion with Newton's identity exposes an exact algebraic structure underlying the one-dimensional finite-bead partition function. The same recursion governs the $d=2$ case of Eq.~\eqref{mainRecursion}, with the cycle traces replaced by $(z_i^N)^2$, and it remains an instance of Newton's identity there. However, the algebraic mapping that closes the one-dimensional case does not carry over, and we have not found a closed form within this framework in higher dimensions. Hence, solving the $d\ge2$ case might require more abstraction or reformulation into another equivalent mathematical definition.

\section{Analytical evaluation of energy and specific heat}
\label{sec6}

\begin{figure}[h]
    \centering
    \includegraphics[width=0.5\linewidth]{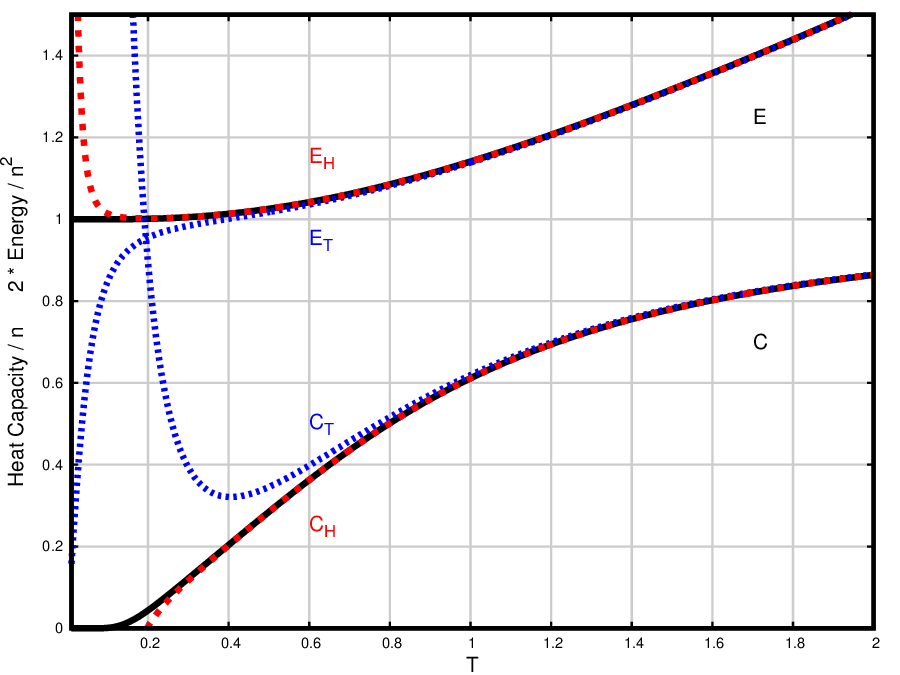}
    \caption{Heat capacity and Energy estimators using analytical results in section \ref{sec6}. Black solid curves show the continuum limit, while the thick dashed red (Hamiltonian) and thin dashed blue (Thermodynamic) curves correspond to the PA propagator with $N=8$. Correct asymptotic behavior is observed for the energy (linear) and heat capacity (one degree of freedom per particle). Results agree with calculations by Takahashi and Imada\cite{tak84a}.}
    \label{fig:CandE}
\end{figure}

Using the closed form for the $d=1$ partition function, the Thermodynamic energy can be computed from
\begin{equation}
\label{eq:Efone}
\begin{split}
     E_n^{T,N} &= -\frac{d}{d\tau}\log(Z_n^{N}) 
     = -\frac{db}{d\tau}\frac{d}{db}\left(\frac{n^2}{2}\log(b) - 
\sum_{k=1}^n \log(1-b^k)\right), \\
     &= \frac{du}{d\epsilon} \left(\frac{n^2}{2} + \sum_{k=1}^n \frac{kb^k}{1-b^k}\right).
\end{split}
\end{equation}
Similarly, using $\tau = \frac{1}{T}$ the corresponding 
specific heat can be computed as
\begin{equation}
    C^{T,N}_n =\frac{\partial E^{T,N}_n}{\partial T} = T^{-2} \left( \left(\frac{du}{d\epsilon}\right)^2\left( \sum_{k=1}^n \frac{k^2b^k}{(1-b^k)^2}\right) - \frac{1}{N} \frac{d^2u}{d\epsilon^2} \left(\frac{n^2}{2} +\sum_{k=1}^n \frac{kb^k}{1-b^k} \right)\right).
\end{equation}
The Hamiltonian energy can be obtained by modifying the prefactor of equation (\ref{eq:Efone}) as in\cite{10.1063/5.0164086}
\begin{equation}
     \label{Henergy}
     E_n^{H,N} = \frac{1}{2}\left(\frac{1}{\gamma}+\gamma\right) \left(\frac{n^2}{2}\ + \sum_{k=1}^n \frac{kb^k}{1-b^k}\right) ,
\end{equation}
The specific heat corresponding to the Hamiltonian energy is then given by
\begin{equation}
\begin{split}
    C^{H,N}_n &=\frac{\partial E^{H,N}_n}{\partial T}\\ &= \frac{T^{-2}}{2} \left( \frac{du}{d\epsilon}\left(\frac{1}{\gamma} + \gamma \right)\left( \sum_{k=1}^n \frac{k^2b^k}{(1-b^k)^2}\right) - \frac{1}{N} \frac{d}{d\epsilon}\left( \frac{1}{\gamma} + \gamma \right) \left(\frac{n^2}{2} +\sum_{k=1}^n \frac{kb^k}{1-b^k} \right)\right).
\end{split}
\end{equation}
Fig. \ref{fig:CandE} illustrates the analytical energies and specific heats for $n=4$ fermions that reproduce the original calculation of Takahashi and Imada \cite{tak84a}.
In the continuum limit $(u\rightarrow\epsilon$, $\gamma\rightarrow 1$, $b\rightarrow {\rm e}^{-\tau})$, and as $\tau \rightarrow \infty$ both energies (\ref{eq:Efone}) and (\ref{Henergy}) converge to the correct 1d fermion energy
\begin{equation*}
     E_n^H = \frac{n^2}{2}.
\end{equation*}
\section{Exact scalability and propagator convergence}
While the generalized finite-bead recursion relation (\ref{mainRecursion}) is formally exact for any dimension $d$, its evaluation in $d=1$ provides a rigorous diagnostic baseline for quantum Monte Carlo methods. In higher dimensions, the fermion sign problem introduces significant statistical noise that obscures the systematic errors arising from finite temporal discretization ($\epsilon$). Utilizing the exact one-dimensional closed-form solution isolates the finite-bead convergence behavior of different short-time propagators across large system sizes without the interference of the sign problem. 

\begin{figure}[H]
\minipage{0.5\textwidth}
  \includegraphics[width=\linewidth]{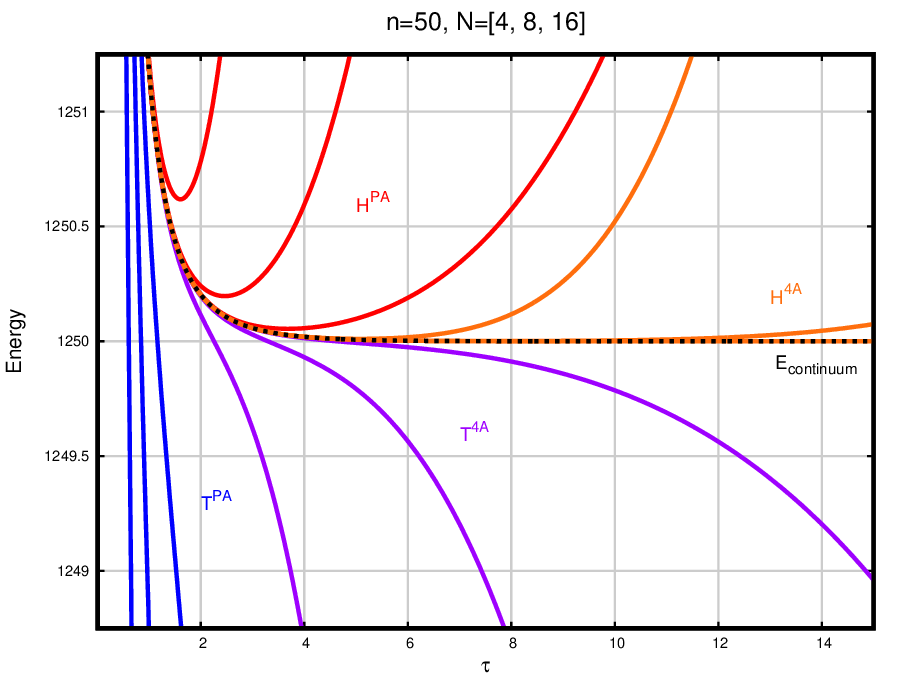}
\endminipage\hfill
\minipage{0.5\textwidth}
  \includegraphics[width=\linewidth]{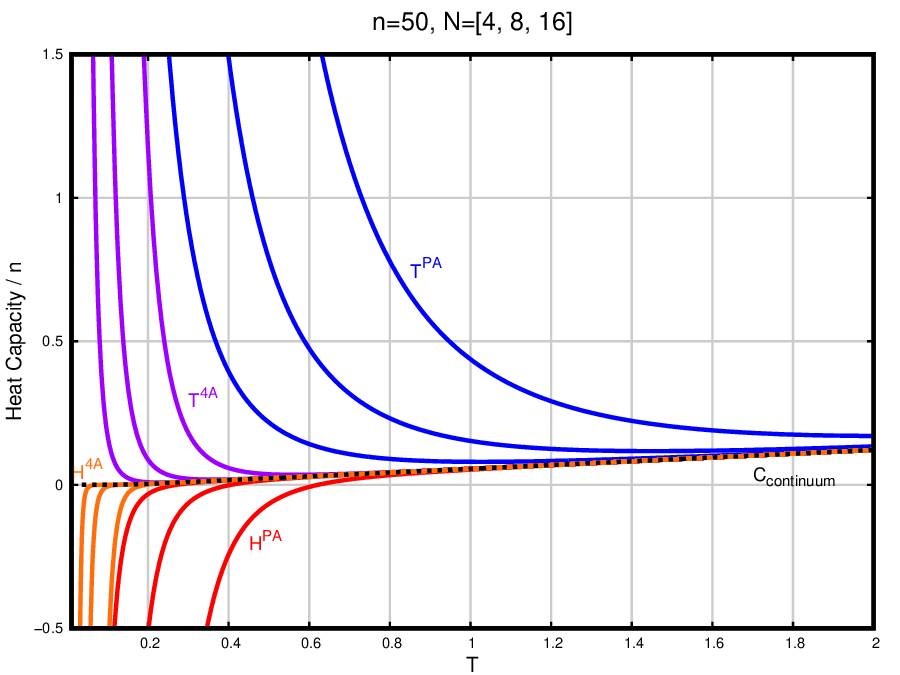}
\endminipage
\caption{Analytical convergence comparison between PA and $4$A short-time propagators for $50$ fermions in one-dimensional harmonic trap. The Hamiltonian energy (left) and specific heat (right) for PA and 4A propagators are shown by red and orange solid lines and labeled $H^{\text{PA}}$ and $H^{4\text{A}}$ respectively. Similarly the thermodynamic energy (left) and specific heat (right) for PA and 4A propagators are shown by blue and purple solid lines and labeled $T^{\text{PA}}$ and $T^{4\text{A}}$ respectively. The continuum limit is shown in the dashed black line, whereas the $4$, $8$ and $16$ bead results are shown left to right for the energy and from right to left for the heat capacity, as convergence increases.}
\label{Fig:lown}
\end{figure}

\begin{figure}[h]
\minipage{0.5\textwidth}
  \includegraphics[width=\linewidth]{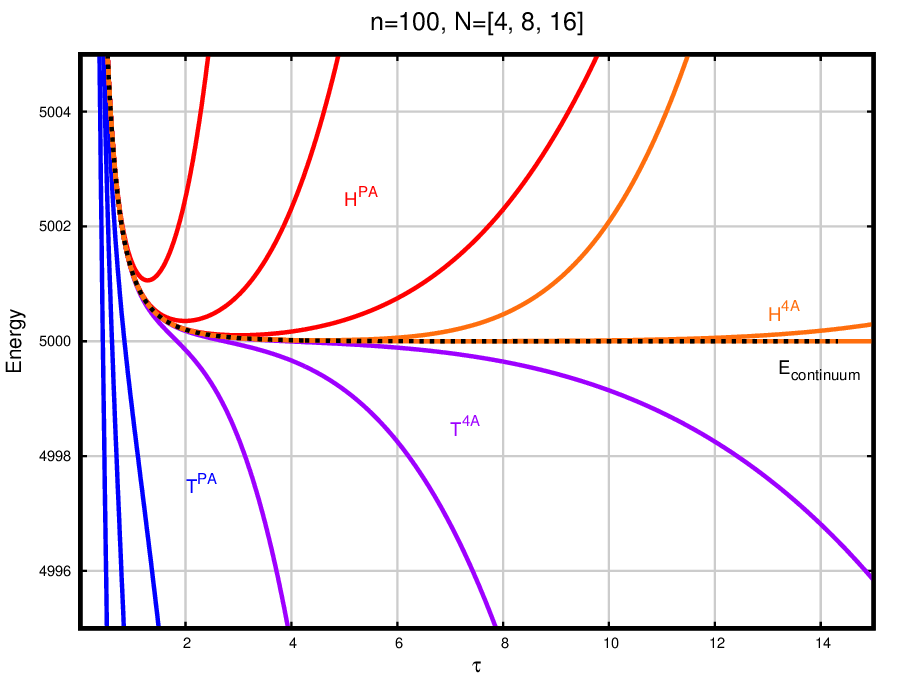}
\endminipage\hfill
\minipage{0.5\textwidth}
  \includegraphics[width=\linewidth]{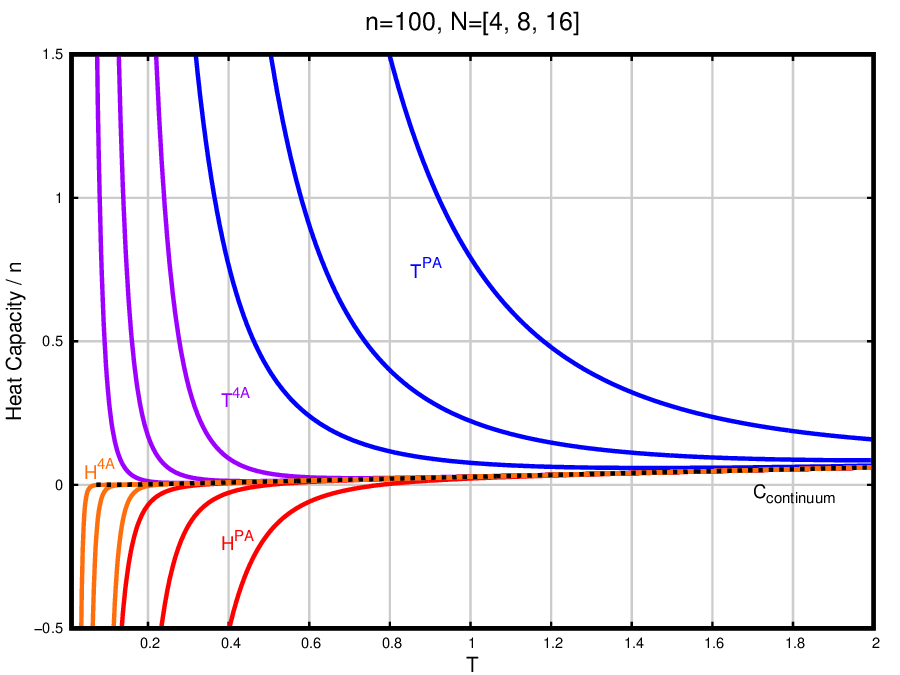}
\endminipage
\caption{Same as Fig. \ref{Fig:lown} for $n=100$ fermions.}
\label{Fig:highn}
\end{figure}

Crucially, the closed form permits exact, scalable evaluation at arbitrary $n$, bypassing the computational scaling limitations associated with numerical PIMC. The ensuing propagator comparison demonstrates the analytical capability of this exact formula, rather than serving purely as a study of one-dimensional propagator physics. In Fig. \ref{Fig:lown} and Fig. \ref{Fig:highn}, this exact scalability is exploited to compare the Thermodynamic and Hamiltonian convergence of the PA and 4A propagators for large systems of $n=50$ and $n=100$ fermions.

When normalized by relative scaling ($0.1\%$ of the exact energy), the high-$\tau$ convergence of the energy for different propagators and bead numbers remains invariant across system sizes. In contrast, the specific heat convergence strictly worsens as the system size increases. %This observation regarding the divergent scaling behavior of the specific heat and energy convergence is a novel finding directly enabled by the exact closed form. 

\section{Conclusion and future directions}
We have derived an exact recursion relation for the finite-bead $n$-fermion partition function in a harmonic trap, valid for arbitrary bead number $N$, imaginary-time $\tau$, and choice of short-time propagator. Unlike continuum recursions, this relation applies directly to the discretized path integral sampled in finite-bead PIMC. By identifying the recursion as an instance of Newton's identity, we used a closed-form result due to Macdonald to obtain an exact expression for the one-dimensional finite-bead partition function. From this expression, the thermodynamic and Hamiltonian energies, as well as the corresponding specific heats, follow analytically, allowing direct comparison of different short-time propagators without the need for cumbersome explicit PIMC calculations.

A natural next step is the extension of this construction to higher dimensions, particularly $d=2$. Since fermions in one-dimension do not exhibit a sign problem, the one-dimensional result provides an analytically controlled foundation but does not yet address the central numerical difficulty of fermionic PIMC. The two-dimensional harmonic trap is therefore the first setting in which the finite-bead recursive framework can be used to study how the sign problem appears in this formulation.

More broadly, the recursive structure of the PIMC integrand derived in Appendix B holds for arbitrary interactions. This suggests that the finite-bead symmetric-function structure may be useful beyond the non-interacting harmonic problem. In particular, organizing the path-integral contribution by cycle structure or representative exchange classes may provide a way to combine permutation sectors analytically before sampling, potentially reducing sign cancellations in interacting fermion systems.

\bigskip
A.C. conceptualized the work and methodology, J.V. developed the software and generated the visualization. Both authors contributed towards formal analysis, wrote, reviewed, and edited the final manuscript. 
\bigskip

\section*{DATA AVAILABILITY}

No data was created or analyzed in this article.

\appendix
\section{Integral of $z_n^N$}
\label{App:Tridiag}
To evaluate this integral, we switch the form of the integral to the following:
\begin{equation}
   z_n^N = \frac{1}{\sqrt{(2\pi\kappa_N(\tau))^n}}\prod_{i=1}^{n} \int dx_i e^{-\frac{1}{2} \sum_{k,l}M_{kl}x_k x_l} 
   = \frac{1}{\sqrt{(2\pi\kappa_N(\tau))^n}} \sqrt{\frac{(2 \pi)^n}{\det(M)}}.
   \label{eq:Znform}
\end{equation}
Here we define $\det(M)$ as
\begin{equation*}
    \det(M) = 
    \begin{vmatrix}
    2\alpha & -\beta  & \cdots & -\beta \\
    -\beta & 2\alpha  & \cdots & 0 \\
    \vdots & \vdots & 
\ddots & \vdots  \\
    -\beta & 0 & \cdots & 2\alpha 
    \end{vmatrix} ^{n \times n}
    \,,\,\,\, \alpha = (\mu_N + \frac{1}{\kappa_N}), \, \beta = \frac{1}{\kappa_N}.
\end{equation*}
In order to take this determinant, we use the tri-diagonal (with corners) determinant identity \cite{TriDiag}:
\begin{equation*}
    \begin{vmatrix}
    a_1 & b_1  & \cdots & c_n \\
    c_1 & a_2  & \cdots & 0 \\
    \vdots & \vdots & \ddots & \vdots  \\
    b_n & 0 & \cdots & a_n 

    \end{vmatrix} ^{n \times n}
    = (-1)^{n+1}(b_n \cdots b_1+c_n \cdots c_1) + \tr \left[
    \begin{pmatrix}
    a_n & -b_{n-1}c_{n-1}\\
    1 & 0\\
    \end{pmatrix}
    \cdots
    \begin{pmatrix}
    a_2 & -b_{1}c_{1}\\
    1 & 0\\
    \end{pmatrix}
    \begin{pmatrix}
    a_1 & -b_{n}c_{n}\\
    1 & 0\\
    \end{pmatrix}
    \right].
\end{equation*}
Applying this identity to $\det(M)$ gives
\begin{equation*}
    \det(M)
    = 2(-1)^{2n+1}\beta^n + \tr\left[
  \begin{pmatrix}
    2\alpha & -\beta^2\\
    1 & 0\\
    \end{pmatrix}^n
    \right]
    = -2\beta^n + \tr\left[A^n\right].
\end{equation*}
The trace can be computed by diagonalizing the matrix A.
\begin{equation*}
    \tr\left[A^n\right]
    = \tr[SD^nS^{-1}] = \tr[D^n] = \tr\left[
    \begin{pmatrix}
    (\alpha-\sqrt{\alpha^2-\beta^2})^n & 0\\
    0 & (\alpha+\sqrt{\alpha^2-\beta^2})^n\\
    \end{pmatrix}
    \right].
\end{equation*}
Hence we get that
\begin{equation*}
    \det(M)
    = (\alpha-\sqrt{\alpha^2-\beta^2})^n + (\alpha+\sqrt{\alpha^2-\beta^2})^n - 2\beta^n.
\end{equation*}
Plugging the determinant back into Eq. (\ref{eq:Znform}) and using the definitions of $\alpha$, $\beta$, and $\mu_N\kappa_N+1 = \zeta_N$ we get
\begin{equation*}
   z_n^N = \sqrt{\frac{1}{(\zeta_N-\sqrt{\zeta_N^2-1})^n + (\zeta_N+\sqrt{\zeta_N^2-1})^n - 2}}.
\end{equation*}
Furthermore, since $\zeta_N = 2\sinh^2(Nu/2) +1$, we arrive at an elementary result for $z_n$,
\begin{equation*}
    z_n^N = \frac{1}{2\sinh(nNu/2)}= \frac{b^{n/2}}{1-b^{n}}.
\end{equation*}
where $b = e^{-Nu}$.

\section{Analysis of multi-bead propagator with arbitrary interactions}
\label{App:Interactions}

The contracted coefficients derived by Chin\cite{10.1063/5.0164086} for the harmonic potential are not applicable to arbitrary potentials, let alone when interactions are included.
This means that we must re-derive this result for multi-bead propagators. While we cannot get a recursive result in terms of $Z_n$, we can show that the integrand can be manipulated to give the same Newton identity structure that integrates to the same value as the $\det$.

We first consider the two-bead case:
\begin{equation*}
\begin{split}
    Z_n(\tau) &= \frac{1}{n!}\prod_{i=1}^{n} \int dx_i dx'_i G(x,x',\tau) G(x',x,\tau) \\
    &= \frac{1}{n!}\frac{1}{(2\pi\kappa(\tau))^{nd}}\prod_{i=1}^{n} \int dx_i dx'_i e^{-2\mu(\tau)  V(x_i)}e^{-2\mu(\tau)  V(x_i')} \times \\ &\qquad \prod_{j=1}^{i-1} e^{-2\mu(\tau)(V(|x_i-x_j|))} e^{-2\mu(\tau)(V(|x_i'-x_j'|))} \det(K(x,x')) \det(K(x',x)).
\end{split}
\end{equation*}
We first simplify by replacing the two determinants with one using $\det(A)\det(B) = \det(AB)$. The product matrix, denoted $K^{(1)}$, has components of the following form:
\begin{equation*}
\begin{split}
   K^{(1)}_{ij}(x,x') &\equiv K_{ik}(x,x') K_{kj}(x',x) = \sum_k e^{-\frac{1}{2\kappa}(x_i - x'_k)^2}e^{-\frac{1}{2\kappa}(x_k' - x_j)^2} \\
   &= \sum_k \exp\!\left(-\frac{1}{2\kappa}(x_i^2 + x_j^2 + 2x'_k{}^2 - 2(x_i+x_j)x'_k)\right).
\end{split}
\end{equation*}
In the two-bead case, $K^{(1)}$ is manifestly symmetric, however, this property does not generally hold for $N$ beads.
We similarly define the $N$-product $K^{(N)}$ and its components as
\begin{equation*}
\begin{split}
    K^{(N)}_{ij} &\equiv K_{i,k_1}(x,x^{(1)}) K_{k_1,k_2}(x^{(1)},x^{(2)}) \cdots K_{k_N,j}(x^{(N)},x) \\
    &= \sum_{k_1,\cdots,k_N} \exp\!\left(-\frac{1}{2\kappa} \left(x_i^2 + x_j^2 + 2\sum_{l=1}^N \bigl(x_{k_l}^{(l)}\bigr)^2 - 2x_i x^{(1)}_{k_1} - 2x^{(1)}_{k_1} x^{(2)}_{k_2} \cdots - 2 x^{(N)}_{k_N} x_j\right)\right).
\end{split}
\end{equation*}
For brevity, explicit dependence on these variables is omitted in subsequent equations.
Having reduced the expression to a single determinant, we proceed with an analysis analogous to that in Section \ref{rfpf}.
With the coefficients $1/{n!} \cdot (2\pi\kappa(\tau))^{-\frac{n(N+1)d}{2}}$ ignored, the partition function for the $(N+1)$-bead case then takes the form
\begin{equation*}
\begin{split}
    Z_n(\tau) &= \int dx_1\cdots dx^{(N)}_n \prod_{i=1}^{n} \vartheta_i(x_1, \cdots, x_i) \prod_{l=1}^N \vartheta_i(x^{(l)}_1, \cdots, x^{(l)}_i) \det(K^{(N)}). \\
    &= \int dx_1\cdots dx^{(N)}_n P \prod_{l=1}^N P^{(l)} \det(K^{(N)})
\end{split}
\end{equation*}
Here we have defined $\vartheta_{i}(x^{(l)}_1, \cdots, x^{(l)}_i) \equiv e^{-2\mu V(x_i^{(l)})} \prod_{j=1}^{i-1} e^{-2\mu V(|x_i^{(l)}-x_j^{(l)}|)}$, which encapsulates the interaction of the $i^{th}$ particle with all the previous particles and the 
external potential. We have also defined $P^{(l)} = \prod_{i=1}^n \vartheta_i(x^{(l)}_1, \cdots, x^{(l)}_i)$ for compactness.
Since the product of these $P^{(l)}$'s stay together and does not depend on $\{x_1,\cdots,x_n\}$, we define $P^N_n \equiv \prod_{l=1}^N P^{(l)}$.
We again evaluate the determinant using the last row and first look at the term with the $K^{(N)}_{nn}$ factor.
\begin{equation*}
\begin{split}
       &\int dx_1\cdots dx^{(N)}_n\, P P^N_n\, K^{(N)}_{nn}
    \begin{vmatrix}
    K^{(N)}_{11} & K^{(N)}_{12} & \cdots & K^{(N)}_{1,n-1} \\
    K^{(N)}_{21} & K^{(N)}_{22} & \cdots & K^{(N)}_{2,n-1} \\
    \vdots  & \vdots  & \ddots & \vdots  \\
    K^{(N)}_{n-1,1} & K^{(N)}_{n-1,2} & \cdots & K^{(N)}_{n-1,n-1}
    \end{vmatrix} \\
    &= \int dx_1\cdots dx^{(N)}_n \left(\int dx_n\,\vartheta_n\, K^{(N)}_{nn}\right) \prod_{i=1}^{n-1} \vartheta_i\, P^N_n\, |K_{n-1}^{(N)}|.
\end{split}
\end{equation*}
To establish a recursive sequence analogous to the non-interacting case, we systematically track the integrated variables.
We define the integrand of the base term $W_{n-1}$, the extracted factor $w^n_1$, and its integrated form $\Tilde{w}^n_1$ as:
\begin{equation*}
\begin{split}
    w^n_1(x_1,\cdots, x^{(N)}_n) &\equiv \vartheta_n\, K^{(N)}_{nn} \\
    W_{n-1}(x_1,\cdots,x_{n-1}; x_1^{(1)},\cdots, x^{(N)}_n) &\equiv |K_{n-1}^{(N)}|\, P^N_n \prod_{i=1}^{n-1} \vartheta_i \\
    \Tilde{w}^n_1(x_1,\cdots,x_{n-1}; x_1^{(1)},\cdots, x^{(N)}_n) &= \int dx_n w^n_1(x_1,\cdots, x^{(N)}_n).
\end{split}
\end{equation*}
The integral for the first term of the determinant expansion then becomes $\int dx_1 \cdots dx_{n-1} \Tilde{w}^n_1 W_{n-1}$.
To evaluate the remaining terms of the determinant, we expand along the last row.
The cofactor sign of the $(n,k)$ entry is $(-1)^{n+k}$. Row $k$, which lacks a diagonal element in the remaining minor, is then moved to the $(n-1)$-th position by $n-1-k$ row swaps, contributing $(-1)^{n-1-k}$. The combined sign is $(-1)^{(n+k)+(n-1-k)} = (-1)^{2n-1} = -1$ for all $k$, exactly as in Section \ref{rfpf}.

Since the indices $1$ through $n-1$ are dummy integration variables, we can cyclically relabel them.
Crucially, the interaction functions $P P^N_n$ are invariant under any permutation of these particle labels.
Therefore, the integrands for all off-diagonal terms are identical after relabeling, yielding $(n-1)$ equivalent contributions:
\begin{equation*}
    Z_n = \int dx_1\cdots dx_{n-1} \Tilde{w}^n_1 W_{n-1} - (n-1) \int d^nx P_n^N \prod_{i=1}^{n} \vartheta_i \, K^{(N)}_{n,n-1} 
    \begin{vmatrix} 
    K^{(N)}_{11} & \cdots & K^{(N)}_{1,n-2} & K^{(N)}_{1,n} \\ 
    \vdots & \ddots & \vdots & \vdots \\ 
    K^{(N)}_{n-2,1} & \cdots & K^{(N)}_{n-2,n-2} & K^{(N)}_{n-2,n} \\ 
    K^{(N)}_{n-1,1} & \cdots & K^{(N)}_{n-1,n-2} & K^{(N)}_{n-1,n} 
    \end{vmatrix}.
\end{equation*}
The inner determinant in the second term can be decomposed in the exact same manner by expanding along its new last row (the $(n-1)$-th row).
Extracting $K^{(N)}_{n-1,n-2}$ yields $(n-2)$ identical terms:
\begin{equation*}
\begin{split} 
    Z_n &= \int dx_1\cdots dx_{n-1} \Tilde{w}^n_1 W_{n-1} - (n-1)\Bigg(\int dx_1\cdots dx_{n-2} \Tilde{w}^n_2 W_{n-2} - \\ 
    &(n-2) \int d^nx P_n^N \prod_{i=1}^{n} \vartheta_i \, K^{(N)}_{n,n-1} K^{(N)}_{n-1,n-2} 
    \begin{vmatrix} 
    K^{(N)}_{11} & \cdots & K^{(N)}_{1,n-3} & K^{(N)}_{1,n} \\ 
    \vdots & \ddots & \vdots & \vdots \\ 
    K^{(N)}_{n-3,1} & \cdots & K^{(N)}_{n-3,n-3} & K^{(N)}_{n-3,n} \\ 
    K^{(N)}_{n-2,1} & \cdots & K^{(N)}_{n-2,n-3} & K^{(N)}_{n-2,n} 
    \end{vmatrix} 
    \Bigg).
\end{split}
\end{equation*}
Iterating this procedure fully expands the determinant and establishes the generalized recursive sequence.
Restoring the coefficient $1/n!$, we define the generalized $i$-th extracted sequence $w^n_i$ and the remaining minor $W_{n-i}$ as:
\begin{gather*} 
    w^n_i(x_1,\cdots, x^{(N)}_n) \equiv K^{(N)}_{n,n-1}\, K^{(N)}_{n-1,n-2} \cdots K^{(N)}_{n-i+2,n-i+1}\, K^{(N)}_{n-i+1,n} \prod_{j=n-i+1}^n \vartheta_{j},\\ 
    W_{n-i}(x_1,\cdots,x_{n-i}; x_1^{(1)},\cdots, x^{(N)}_n) \equiv \frac{1}{(n-i)!}|K_{n-i}^{(N)}|\, P_n^N \prod_{j=1}^{n-i} \vartheta_{j}. \\
    W_n \equiv \frac{1}{n} \sum_{i=1}^n (-1)^{i-1} w^n_i W_{n-i}.
\end{gather*}
We can separate the kinetic and potential terms by noticing that $P P_n^N$ is present in all terms.
We then have
\begin{equation*}
    W_n = \frac{1}{n} P P_n^N \sum_{i=1}^n (-1)^{i-1} \eta_i^n H_{n-i},
\end{equation*}
where
\begin{equation*}
\begin{split}
    \eta_i^n &\equiv K^{(N)}_{n,n-1}\, K^{(N)}_{n-1,n-2} \cdots K^{(N)}_{n-i+2,n-i+1}\, K^{(N)}_{n-i+1,n} \\
    H_{n-i} &\equiv \frac{1}{(n-i)!}|K_{n-i}^{(N)}|.
\end{split}
\end{equation*}
Restoring the coefficient $(2\pi\kappa(\tau))^{-\frac{n(N+1)d}{2}}$, this yields a recursion structurally identical to the previous relation, albeit with generalized definitions:
\begin{equation}
\begin{split}
    W_n &= H_n P P_n^N \\
    H_n &= \frac{1}{n} \sum_{i=1}^n (-1)^{i-1} \eta^n_i H_{n-i} \\
    Z_n &= \frac{1}{(\sqrt{2\pi\kappa(\tau)}\,)^{n(N+1)d}} \int dx_1 \cdots dx_n^{(N)}\, \, W_n.
\end{split}
\end{equation}

This recursive relation here is equivalent to the $\det$ definition only after integration. Hence, what we showed here is that after integration over permutation-invariant variables, the off-diagonal cofactor contributions can be grouped into equivalent classes and result in a recursive definition. Furthermore, compared to the determinant definition with $\sim n!$ terms, this definition only has $\sim 2^{n-1}$ terms, evaluable in $O(n^2)$, making separation of different parity permutations much easier.

\bibliography{ref1.1} % Entries are in the refs.bib file

@article{10.1063/1.464180,
    author = {Borrmann, Peter and Franke, Gert},
    title = {Recursion formulas for quantum statistical partition functions},
    journal = {J. Chem. Phys.},
    volume = {98},
    pages = {2484-2485},
    year = {1993},
    doi = {10.1063/1.464180},
    url = {https://doi.org/10.1063/1.464180}
}

@article{PhysRevE.55.227,
  title = {Thermodynamics of coupled identical oscillators within the path-integral formalism},
  author = {Brosens, F. and Devreese, J. T. and Lemmens, L. F.},
  journal = {Phys. Rev. E},
  volume = {55},
  pages = {227--236},
  year = {1997},
  doi = {10.1103/PhysRevE.55.227},
  url = {https://link.aps.org/doi/10.1103/PhysRevE.55.227}
}

@article{Schmidt_2002,
   title={Partition functions and symmetric polynomials},
   volume={70},
   url={http://dx.doi.org/10.1119/1.1412643},
   DOI={10.1119/1.1412643},
   journal={Am. J. Phys.},
   author={Schmidt, Heinz-Jürgen and Schnack, Jürgen},
   year={2002},
   pages={53–57} 
}

@book{macdonald1998symmetric,
  title={Symmetric Functions and Hall Polynomials},
  author={Macdonald, I.G.},
  isbn={9780198504504},
  lccn={94027392},
  series={Oxford classic texts in the physical sciences},
  url={https://books.google.com/books?id=srv90XiUbZoC},
  year={1998},
  publisher={Clarendon Press}
}

@article{tak84a,
    author = {Takahashi ,Minoru and Imada ,Masatoshi},
    title = {Monte Carlo Calculation of Quantum Systems},
    journal = {J. Phys. Soc. Jpn.},
    volume = {53},
    pages = {963-974},
    year = {1984}
}

@article{TriDiag,
   title={Determinants of block tridiagonal matrices},
   volume={429},
   url={http://dx.doi.org/10.1016/j.laa.2008.06.015},
   DOI={10.1016/j.laa.2008.06.015},
   journal={Linear Algebra Appl.},
   author={Molinari, Luca Guido},
   year={2008},
   pages={2221–2226} 
}

@article{10.1063/5.0164086,
    author = {Chin, Siu A.},
    title = {Anatomy of path integral Monte Carlo: Algebraic derivation of the harmonic oscillator’s universal discrete imaginary-time propagator and its sequential optimization},
    journal = {J. Chem. Phys.},
    volume = {159},
    pages = {134109},
    year = {2023},
    doi = {10.1063/5.0164086},
    url = {https://doi.org/10.1063/5.0164086}
}

@article{10.1063/1.1485725,
    author = {Chin, Siu A. and Chen, C. R.},
    title = {Gradient symplectic algorithms for solving the Schrödinger equation with time-dependent potentials},
    journal = {J. Chem. Phys.},
    volume = {117},
    pages = {1409-1415},
    year = {2002},
    doi = {10.1063/1.1485725},
    url = {https://doi.org/10.1063/1.1485725}
}

@article{Ford71,
    author = {Ford, D. I.},
    title = {A Note on the Partition Function for Systems of Independent Particles},
    journal = {Am. J. Phys.},
    volume = {39},
    number = {2},
    pages = {215-220},
    year = {1971},
    doi = {10.1119/1.1986094},
    url = {https://doi.org/10.1119/1.1986094}
}

@article{PhysRevE.107.035305,
  title = {No sign problem in one-dimensional path integral Monte Carlo simulation of fermions: A topological proof},
  author = {Chin, Siu A.},
  journal = {Phys. Rev. E},
  volume = {107},
  pages = {035305},
  year = {2023},
  doi = {10.1103/PhysRevE.107.035305},
  url = {https://link.aps.org/doi/10.1103/PhysRevE.107.035305}
}

@article{chin2026understandingsignproblemexact,
    author = {Chin, Siu A.},
    title = {Understanding the sign problem from an exact path integral Monte Carlo model of interacting harmonic fermions},
    journal = {J. Chem. Phys.},
    volume = {164},
    pages = {224122},
    year = {2026},
    doi = {10.1063/5.0324529},
    url = {https://doi.org/10.1063/5.0324529}
}
\end{document}